\documentclass[aps,prd,nofootinbib,showpacs,floatfix,twocolumn,superscriptaddress]{revtex4}

\usepackage{epsfig}
\usepackage{epstopdf}
\usepackage{mathrsfs}
\usepackage{amsmath}
\usepackage{amssymb}
\usepackage{color}
\usepackage{graphicx}
\usepackage{dcolumn}
\usepackage{multirow}
\usepackage{bm}

\newcommand{\be}{\begin{equation}}
\newcommand{\ee}{\end{equation}}
\newcommand{\ba}{\begin{eqnarray}}
\newcommand{\ea}{\end{eqnarray}}

\def\xm{X_{\rm max}}

\begin{document}

\title{Depth of Maximum of Air-Shower Profiles at the Pierre Auger Observatory: Composition Implications}

\author{A.~Aab}
\affiliation{Universit\"{a}t Siegen, Siegen, 
Germany}
\author{P.~Abreu}
\affiliation{Laborat\'{o}rio de Instrumenta\c{c}\~{a}o e F\'{\i}sica 
Experimental de Part\'{\i}culas - LIP and  Instituto Superior 
T\'{e}cnico - IST, Universidade de Lisboa - UL, 
Portugal}
\author{M.~Aglietta}
\affiliation{Osservatorio Astrofisico di Torino  (INAF), 
Universit\`{a} di Torino and Sezione INFN, Torino, 
Italy}
\author{E.J.~Ahn}
\affiliation{Fermilab, Batavia, IL, 
USA}
\author{I.~Al Samarai}
\affiliation{Institut de Physique Nucl\'{e}aire d'Orsay (IPNO), 
Universit\'{e} Paris 11, CNRS-IN2P3, Orsay, 
France}
\author{I.F.M.~Albuquerque}
\affiliation{Universidade de S\~{a}o Paulo, Instituto de F\'{\i}sica, 
S\~{a}o Paulo, SP, 
Brazil}
\author{I.~Allekotte}
\affiliation{Centro At\'{o}mico Bariloche and Instituto Balseiro 
(CNEA-UNCuyo-CONICET), San Carlos de Bariloche, 
Argentina}
\author{J.~Allen}
\affiliation{New York University, New York, NY, 
USA}
\author{P.~Allison}
\affiliation{Ohio State University, Columbus, OH, 
USA}
\author{A.~Almela}
\affiliation{Universidad Tecnol\'{o}gica Nacional - Facultad 
Regional Buenos Aires, Buenos Aires, 
Argentina}
\affiliation{Instituto de Tecnolog\'{\i}as en Detecci\'{o}n y 
Astropart\'{\i}culas (CNEA, CONICET, UNSAM), Buenos Aires, 
Argentina}
% ...
\collaboration{The Pierre Auger Collaboration}
\email{{\tt auger\_spokespersons@fnal.gov}}
\noaffiliation



\begin{abstract}
Using the data taken at the Pierre Auger Observatory between December 2004 and December 2012, we have examined the implications of the distributions of depths of atmospheric shower maximum ($\xm$), using a hybrid technique, for composition and hadronic interaction models. We do this by fitting the distributions with predictions from a variety of hadronic interaction models for variations in the composition of the primary cosmic rays and examining the quality of the fit. Regardless of what interaction model is assumed, we find that our data are not well described by a mix of protons and iron nuclei over most of the energy range. Acceptable fits can be obtained when intermediate masses are included, and when this is done consistent results for the proton and iron-nuclei contributions can be found using the available models. We observe a strong energy dependence of the resulting proton fractions, and find no support from any of the models for a significant contribution from iron nuclei. However, we also observe a significant disagreement between the models with respect to the relative contributions of the intermediate components.
\end{abstract}


\maketitle 

\section{Introduction}

The composition of ultra-high energy cosmic rays (UHECRs) is an important input for elucidating their origin which is yet to be fully understood. The atmospheric depth where the longitudinal development of an air shower reaches the maximum number of particles, $\xm$, is a standard parameter used to extract composition information as different nuclei produce different distributions of $\xm$ \cite{GaisserHillas}. The mean and dispersion of $\xm$ have been previously \cite{Abraham:2010yv} utilized to infer information on the composition, especially since the former scales linearly with the logarithm of the composition mass $\ln A$. Data taken at the Pierre Auger Observatory \cite{Abraham:2004dt} located in Argentina are well suited to study composition as the capabilities of the Observatory for hybrid\footnote{These are events that triggered both the surface and fluorescence detectors. The surface detectors are used to constrain the shower geometry and thereby to reduce the uncertainty in the $\xm$ reconstruction.} detection of air showers enable high-accuracy measurement of the $\xm$ parameter  \cite{longxmax2014}.

\begin{figure*}[t]
\includegraphics[width=0.49\textwidth]{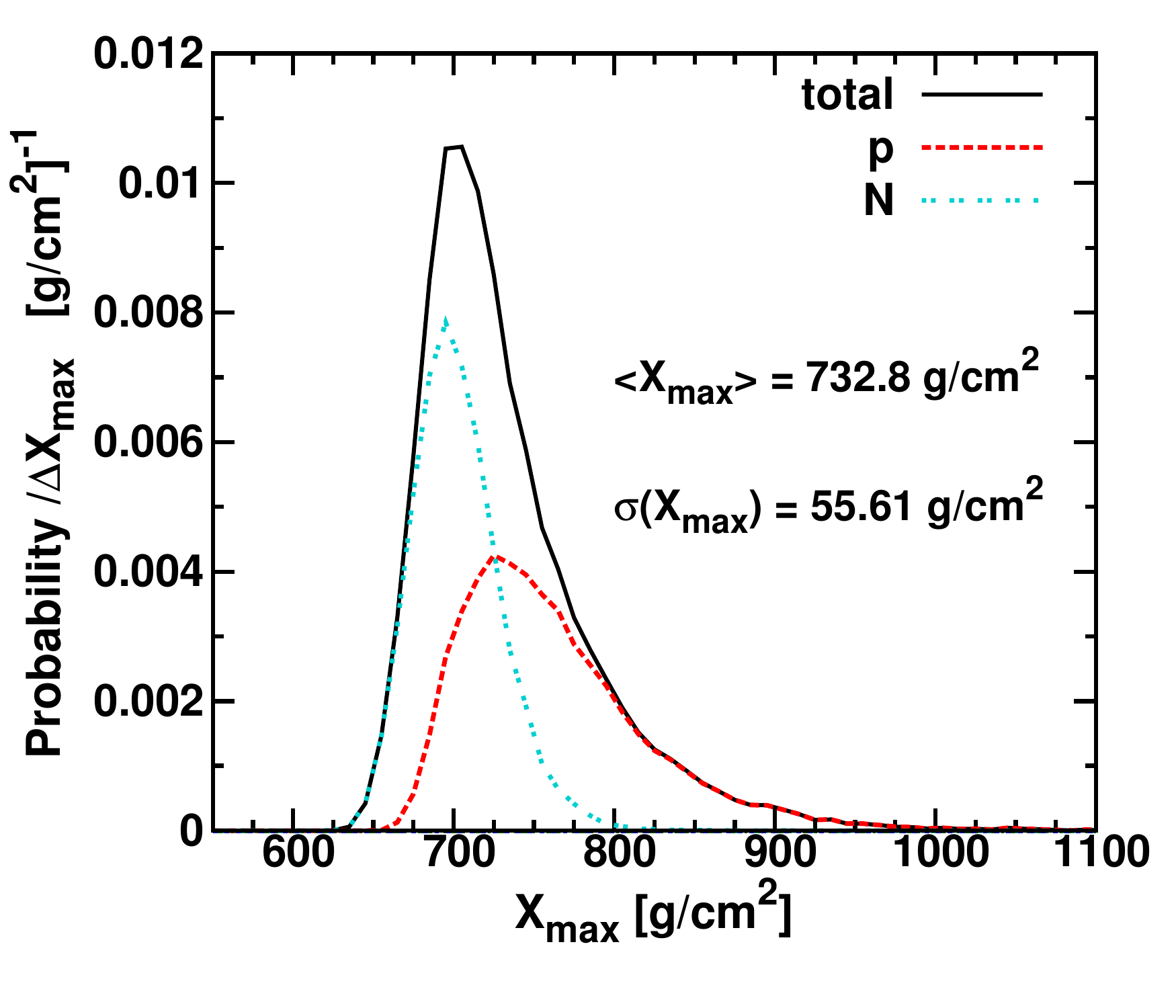} \hspace*{-15mm}
\includegraphics[width=0.49\textwidth]{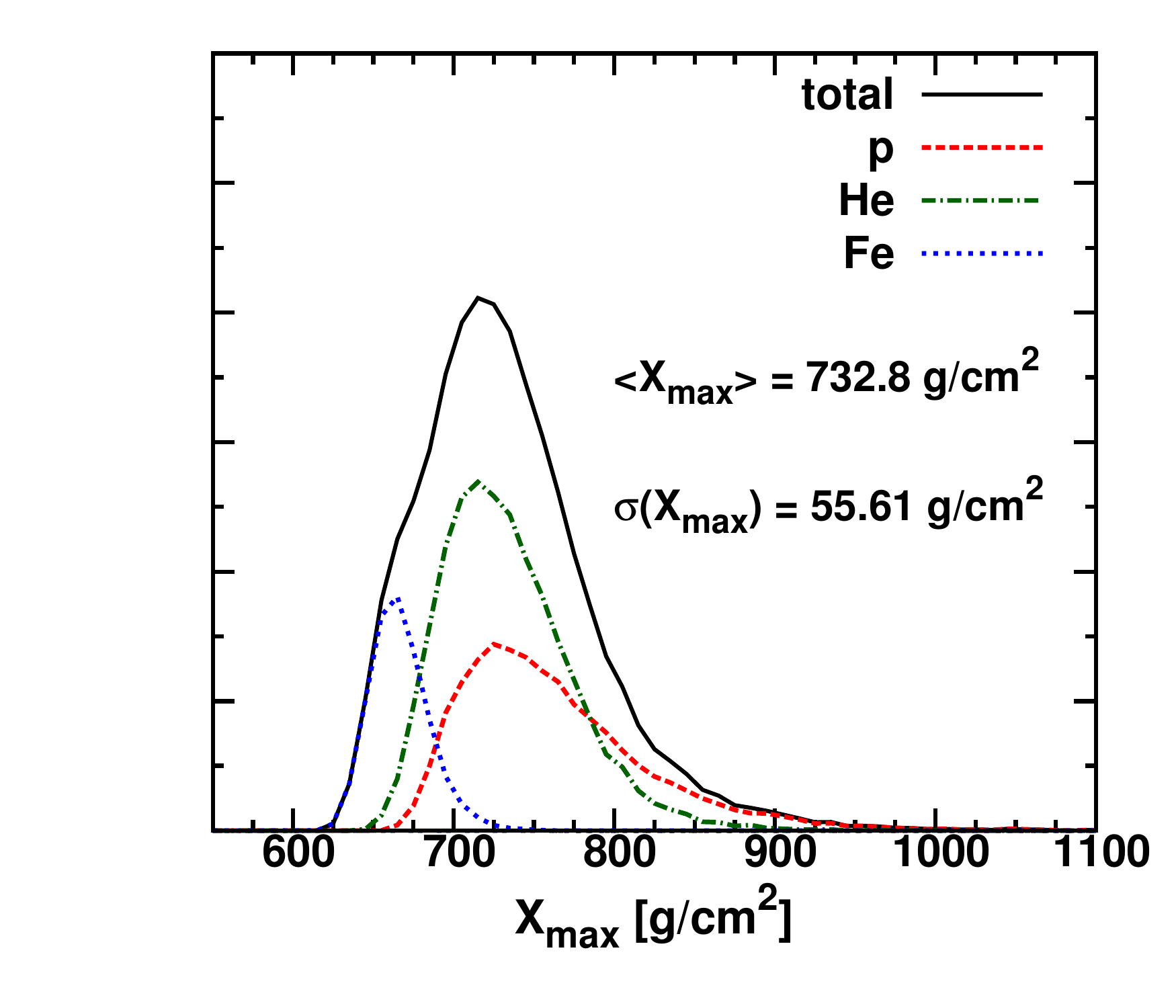}
\vspace*{-5mm}
\caption{Two $X_{\rm max}$ distributions generated with identical mean and dispersion but with different compositions. The hadronic interaction model EPOS-LHC was used to generate $10^4$ events in the range $E=10^{18.2-18.3}$ eV.}
\label{fig_histo}
\end{figure*}

In a recent study \cite{Abreu:2013env}, the mean and dispersion of $\xm$ were converted to the first two moments of the $\ln A$ distribution to deduce the details of the mass composition extracted from the Auger data. That method allowed us to obtain the average logarithmic mass of components that describe the data, as well as testing if that combination is feasible for the given hadronic interaction model used. In this work, we use the shape of the distribution of $\xm$ data from Auger to infer the composition. Using the $\xm$ distribution maximizes the information and helps reduce degeneracies that can occur when one considers only the first two moments of the $X_{\rm max}$ distribution. Figure~\ref{fig_histo} displays two simulated distributions with different mixes of composition but with identical means and dispersions. 
By maintaining sensitivity to the shape of the distribution, information on the composition can be retrieved that goes beyond the mean and dispersion of $\ln A$.

For a given hadronic interaction model, the $\xm$ distribution is compared to predictions made using Monte Carlo (MC) simulations formed with varying nuclear fractions, and a binned maximum-likelihood discriminator is used to choose the best-fit fractions. This method also allows us to obtain information on the goodness of the fit.

The hybrid $\xm$ dataset in the range $E = 10^{17.8} - 10^{20}$ eV measured by Auger \cite{longxmax2014} is used to determine whether it can be described satisfactorily by an evolution of composition with energy. We first consider a mixture of the two most stable types of particles, protons and iron nuclei, and then we extend the fits to include extra components. Specifically, we include helium and nitrogen nuclei as representatives of the intermediate range of nuclear masses.

The procedure used to form the MC predictions is described in Section II, and the fitting procedure is described in Section III. The systematics considered in the analysis are described in Section IV, the results are presented in Section V, and the discussion and conclusions in Section VI.

\section{Templates for Monte Carlo simulations of $\xm$}

A template is developed for the MC simulation of the $\xm$ distribution for a single nuclear species, and is created to compare it with the data. To form a template, we start with the true $\xm$ obtained from events generated in the MC with specific incident species and a given energy range. To generate the simulations, the algorithm CONEX v4r37 \cite{Bergmann:2006yz, Pierog:2004re} has been used to simulate air showers, using the three most common hadronic interaction packages EPOS-LHC \cite{Werner:2005jf}, QGSJet II-4 \cite{Ostapchenko:2005nj} and Sibyll 2.1 \cite{Ahn:2009wx}, where the first two models have been updated with the $E_{CM}=7$ TeV LHC data.

There are $2 \times 10^4$ showers simulated per species per energy bin. The zenith angle is distributed isotropically on a flat surface ($dN/d\cos(\theta) \sim \cos \theta$) between 0 and 80 degrees. The distribution of energy within a given bin follows $E^{-a}$, where $a=1.1$ or 2.2 for energies below or above $10^{18}$ eV, respectively\footnote{This parameterization was derived from the energy distribution of preliminary data, whereas the current dataset is best described by $a=1.76+0.44\log{E/{\rm EeV}}$ \cite{longxmax2014}. The new parameterization would produce at most a 0.3\%  shift in the average energy  within a bin which is negligible when compared to the systematic energy-scale uncertainty.}.

The true $\xm$ for a given nuclear species $s$, $X^{\rm t}_s $, is determined by a quadratic interpolation around the peak as a function of slant depth. The template is a binned $\xm$ distribution that includes effects of acceptance and measurement resolution. The content of the $j$-th bin is the sum of the contributions from the $N_{MC}$ simulated events, each weighted by the acceptance;
\be
X^{\rm m}_{s,j} ~=~ \sum_n^{N_{MC}}a(X^{\rm t}_{s,n}) \, p_j(X^{\rm t}_{s,n}) / N_{MC} ~,
\ee
where $a(X^{\rm t}_{s,n})$ is the acceptance weight for the $n$-th event and $p_j(X^{\rm t}_{s,n})$ is the probability that $\xm$ measured for this event lies within the range defined by the $j$-th bin. This probability is obtained assuming a resolution function represented by a double Gaussian, where the parameters of the dependence on energy have been determined using a full detector simulation \cite{longxmax2014}. Note that $a(X^{\rm t}_{s,n})$ is not included in the normalization of the template so that the sum of $X^{\rm m}_{s,j}$ is somewhat less than 1 by an amount depending on the overall acceptance for a given species arriving within the field of view. This overall factor to correct for acceptance ranges from 0.979 for protons in the EPOS-LHC model, up to 1 for iron nuclei in all models.

\section{Fitting procedure}

We use hybrid data collected with Auger between December 2004 and December 2012, where 19,759 events survived all the cuts with energies of  $E_{\rm lab} = 10^{17.8}$ eV and higher, as described in Ref.~\cite{longxmax2014}. The events are binned in intervals of 0.1 in $\log(E/{\textrm eV})$ from $10^{17.8}$ to $10^{19.5}$ eV and events with energy above $10^{19.5}$ eV are combined into one bin. The number of events ranges from more than 3000 per low-energy bin to about 40 for the highest-energy bin. The $\xm$ bins are defined to be 20 ${\rm g/cm}^2$ wide starting at $\xm=0$. 

To carry out the comparison with data, for a given energy bin the template $X^{\rm m}_{s,j}$ for each species is weighted according to its species fraction $f_s$ and combined to form MC predictions, $C_j$, for each $\xm$ bin:
\be
\displaystyle
C_j ~=~  \frac{N_{data}}{N} \sum_s f_s \, X^{\rm m}_{s,j} ~,
\label{eqn-temp}
\ee
where $N_{data}$ is the number of measured events in the energy bin and the normalization term $N$ is a function of $f_s$
\begin{subequations}
\be
N = \sum_s f_s \sum_j^\infty X^{\rm m}_{s,j} ~,
\ee
with
\be
 \sum_s f_s = 1 \ .
\ee
\end{subequations}
We use the normalizations for the templates and for the predictions to interpret $f_s$ as the fraction of species $s$ at the top of the atmosphere, {\it i.e.}, without the need to correct for detector acceptance.

A binned maximum-likelihood method is used to find the best-fitting combination of the various species. For a given energy bin $E$, the likelihood is expressed as
\be
\displaystyle
L ~=~ \prod_j \left[ \frac{e^{-C_j} C_j^{\,n_j}}{n_j !} \right] \ ,
\ee
where $n_j$ is the measured count of events in $\xm$ bin $j$ and $C_j$ is the corresponding MC prediction. As a practical consideration, we remove the factorials by dividing $L$ by the likelihood value obtained when $C_j=n_j $. As this value is a constant factor, the maximization is not affected by this process. This has the added advantage that the resulting likelihood ratio can also be used as an estimator for the goodness of fit \cite{max1963};
\be
\displaystyle
L^{'} ~=~\prod_j \left[ \frac{e^{-C_j} C_j^{\,n_j}}{n_j !}  \right] / 
\left[ \frac{e^{-n_j} n_j^{\,n_j}}{n_j !}  \right] \ .
\ee
The species fractions $F_i$ that best fit the data are found by minimizing the negative log-likelihood expression
\be
\displaystyle
\mathscr{L}  ~=~ - \ln L^{'} ~=~
 \sum_j \left( C_j \,-\, n_j  \,+\, n_j \ln \frac{n_j}{C_j} \right) \ .
\label{eq-logl}
\ee
The fit quality is measured by the $p$-value which is defined as the probability of obtaining a worse fit (larger $\mathscr{L}$) than that obtained with the data, assuming that the distribution predicted by the fit results is correct. To construct $p$-values for the fit, mock datasets of the predicted $\xm$ distribution were generated from the templates with size equal to the real dataset. The $p$-value was calculated as the fraction of mock datasets with $\mathscr{L}$ worse than that obtained from the real data. Since the parameters in the fit are constrained by both physical and unitarity bounds, we do not expect $\mathscr{L}$ to necessarily behave like a $\chi^2$ variable and hence do not use the $\Delta \mathscr{L} = 1/2$ rule to obtain the statistical uncertainty on the fit parameters. Instead, the statistical uncertainty for each species has been determined by using a generalization of the Feldman-Cousins procedure \cite{Feldman:1997qc}. Known as the profile-likelihood method \cite{Rolke:2004mj}, a multi-dimensional likelihood function is reduced to a function that only depends on the parameter of prime interest. The 68\% confidence range for each species fraction is determined through this method by treating the other species fractions as nuisance parameters. The method properly accounts for correlations and provides a smooth transition from two-sided bounds to one-sided limits.

\section{Systematic uncertainties}

The most important source of systematic uncertainty considered is that on $\xm^{\rm m}$ itself as determined in Ref.~\cite{longxmax2014}. The effect of this uncertainty on the fit fractions is determined by fitting the data with model predictions shifted in $\xm$ by an amount $\delta \xm$. The models are shifted rather than the data in order to avoid statistical artifacts resulting from rebinning of the data. Since we do not expect the fit fractions to evolve monotonically with respect to $\delta \xm$, we scan $\delta \xm$ between $+1 \sigma$ and $-1 \sigma$ in steps of $0.2 \sigma$ in order to determine the maximum range over which a fit fraction can vary.

The other possible systematic uncertainties we considered are those on the energy scale and on the parameterization of the resolution functions for acceptance and $\xm$. The effect of the parameterization uncertainties is evaluated by refitting the data with extreme values of the parameterizations. The latter values were chosen to produce the largest or smallest acceptance or resolution, respectively, compatible with the data \cite{longxmax2014}. None of the parameterization variants resulted in significant changes to the fit fractions. Since the uncertainty in the energy scale is comparable to the width of the energy bin, we evaluated its effect by simply refitting the data with MC templates constructed from adjacent energy bins. The effects on the fit fractions were similar to, but generally smaller than, the shifts in $\xm$ scale. 

The overall systematic uncertainty assigned to a given fit fraction is chosen to encompass the full range of values obtained by any of the fit variants described above. The $p$-values are also calculated for each of these fit variants in order to assess their effect on the goodness of fit.

\section{Results}

\begin{figure*}
\includegraphics[width=0.99\textwidth]{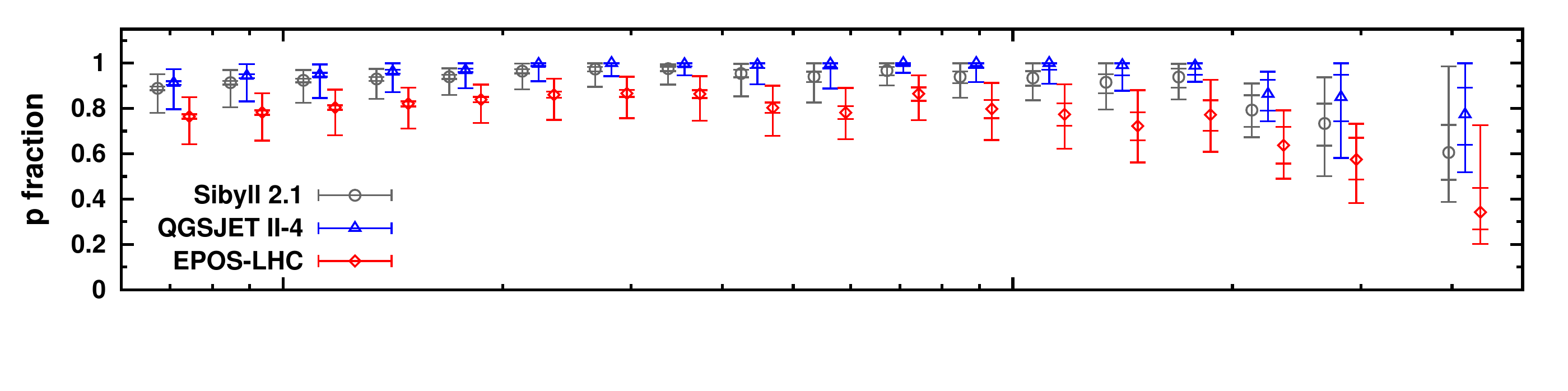}

\vspace*{-12mm}
\includegraphics[width=0.99\textwidth]{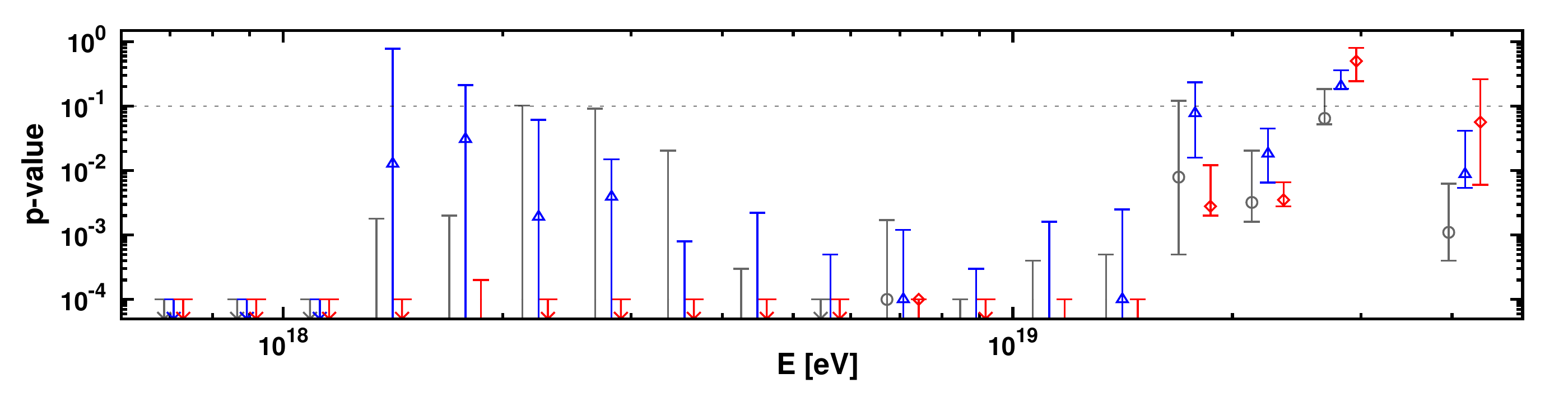}

\vspace*{-5mm}
\caption{Fitted fraction and quality for the scenario with protons and iron nuclei only. The upper panel shows the proton fraction and the lower panel shows the $p$-values. The horizontal dotted line in the lower panel indicates $p=0.1$. The results from the various hadronic interaction models are slightly shifted in energy for better viewing (Sibyll 2.1 to the left, EPOS-LHC to the right).}
\label{fig:two-ori}
\end{figure*}

\begin{figure*}
\includegraphics[width=0.99\textwidth]{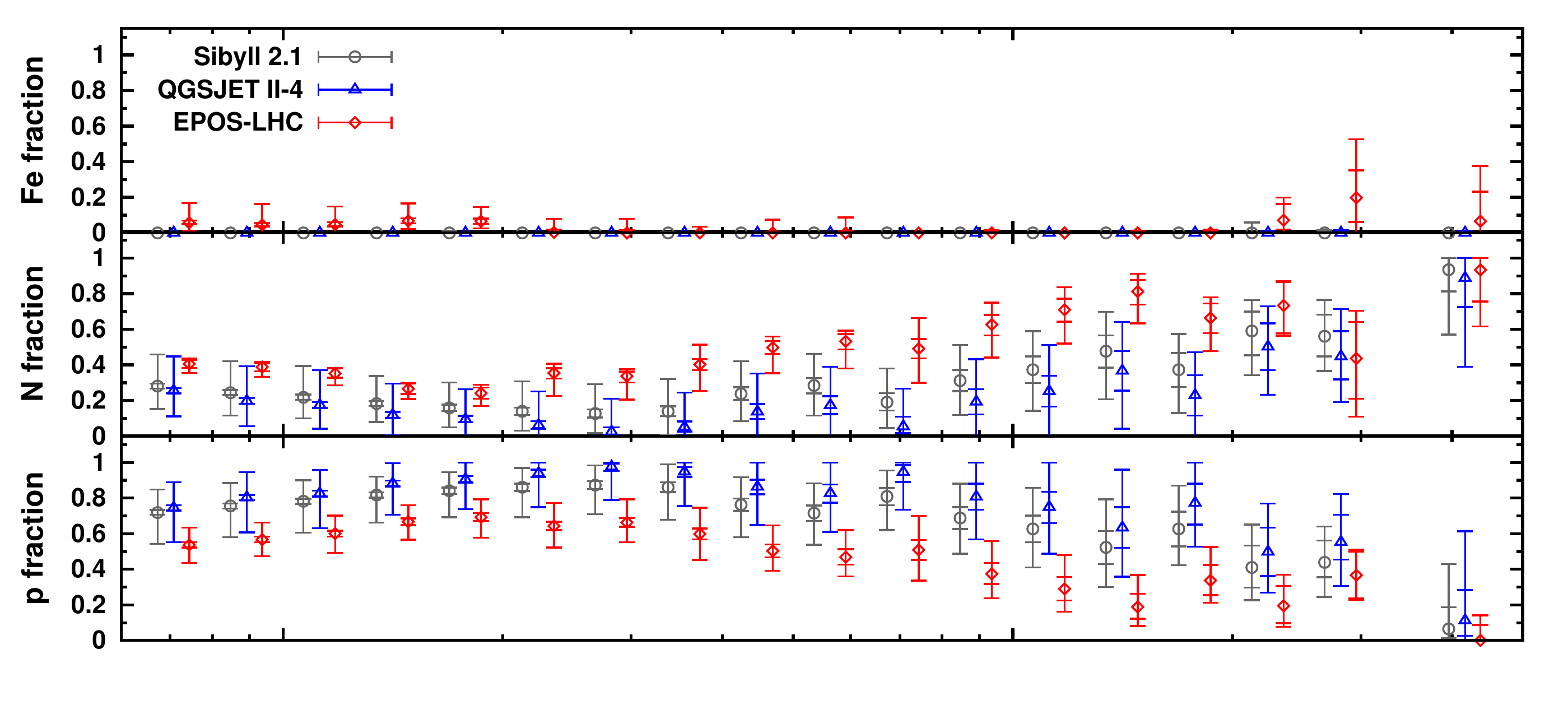}

\vspace*{-10mm}
\includegraphics[width=0.99\textwidth]{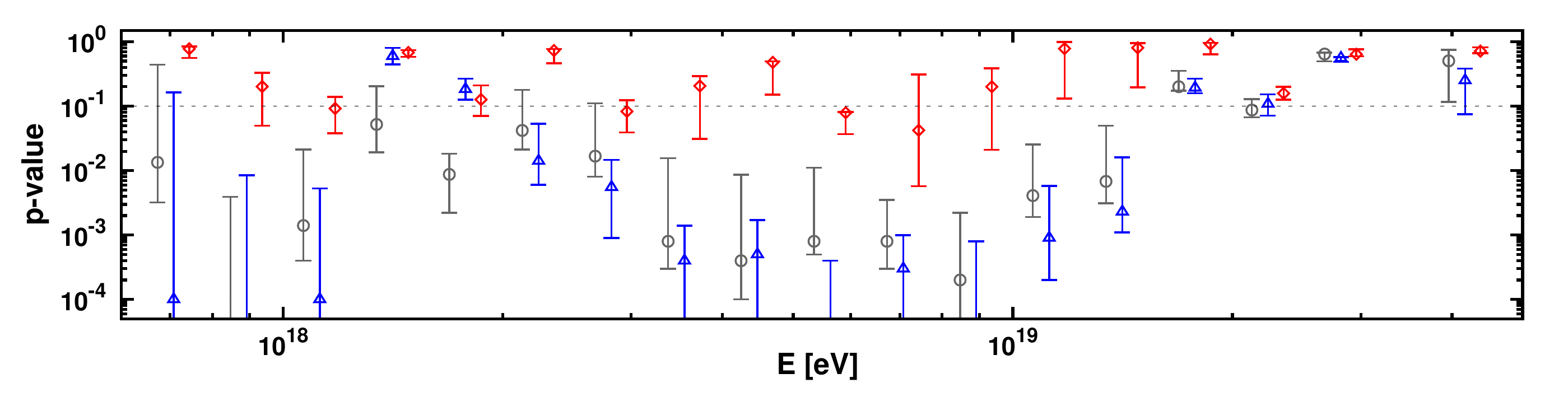}

\vspace*{-5mm}
\caption{Fitted fraction and quality for the scenario of a complex mixture of protons, nitrogen nuclei, and iron nuclei. The upper panels show the species fractions and the lower panel shows the $p$-values.}
\label{fig:three-ori}
\end{figure*}

\begin{figure*}
\includegraphics[width=0.99\textwidth]{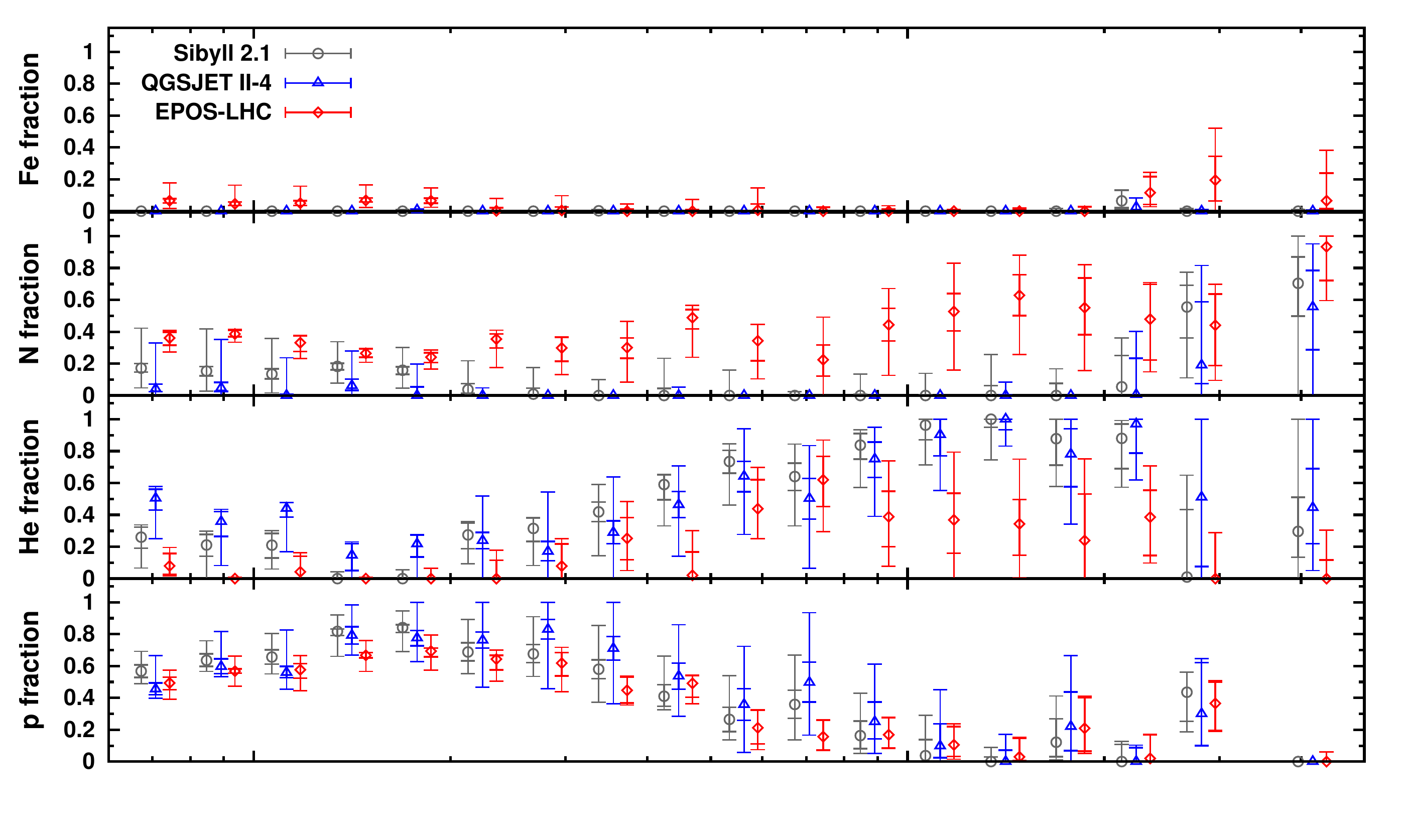}

\vspace*{-10mm}
\includegraphics[width=0.99\textwidth]{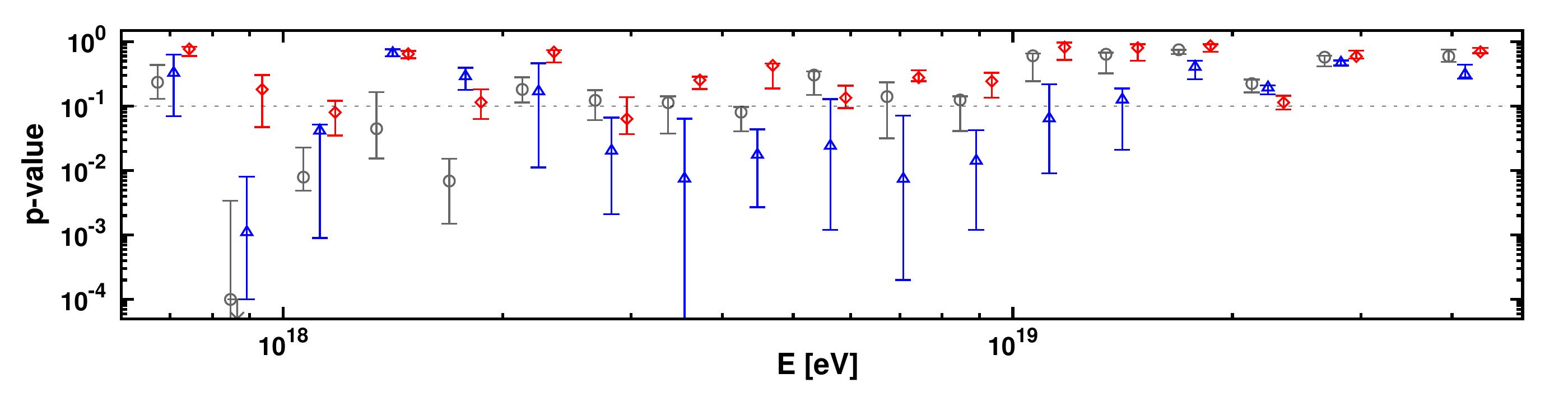}

\vspace*{-5mm}
\caption{Fitted fraction and quality for the scenario of a complex mixture of protons, helium nuclei, nitrogen nuclei, and iron nuclei. The upper panels show the species fractions and the lower panel shows the $p$-values.}
\label{fig:four-ori}
\end{figure*}

The fit result for the mix of protons and iron nuclei is shown in Fig.~\ref{fig:two-ori}. Fit results with additional components are shown in Figs.~\ref{fig:three-ori} and \ref{fig:four-ori}. For each figure the species fractions are shown in the upper panel(s). Only the proton fraction is shown for the combination of protons and iron nuclei (Fig.~\ref{fig:two-ori}), while all species fractions are shown when more than two components are considered (Figs.~\ref{fig:three-ori} and \ref{fig:four-ori}). The inner error bars are statistical and the outer ones include the systematic uncertainty added in quadrature. The $p$-values are shown in the lower panel of the figures, with error bars corresponding to the range of variation obtained within the systematic uncertainties. Where $p$-values are less than $10^{-4}$, they are indicated with downward arrows. 

For the simple mixture of protons and iron nuclei (Fig.~\ref{fig:two-ori}), only the second-highest energy bin ($E = 10^{19.4-19.5}$ eV) yields good fit qualities for all three hadronic interaction models. However the fit qualities for all three models are generally poor throughout the energy range, even when the systematic uncertainties are taken into consideration.

In order to determine whether there is any composition mixture where the models result in an adequate representation of the data, we extended the fits to include extra components. When nitrogen nuclei are added as an intermediate mass term (Fig.~\ref{fig:three-ori}), the quality of the fits is acceptable for EPOS-LHC. However, though much improved, the quality of the fits is still poor over most of the energy range for the other two models. $p$-values for all models are good for events with energy above $10^{19.2}$ eV. When helium nuclei are also included, we find that the data are well described by all models within systematic uncertainties over most of the energy range (Fig.~\ref{fig:four-ori}). 

\begin{figure*}
\includegraphics[width=0.99\textwidth]{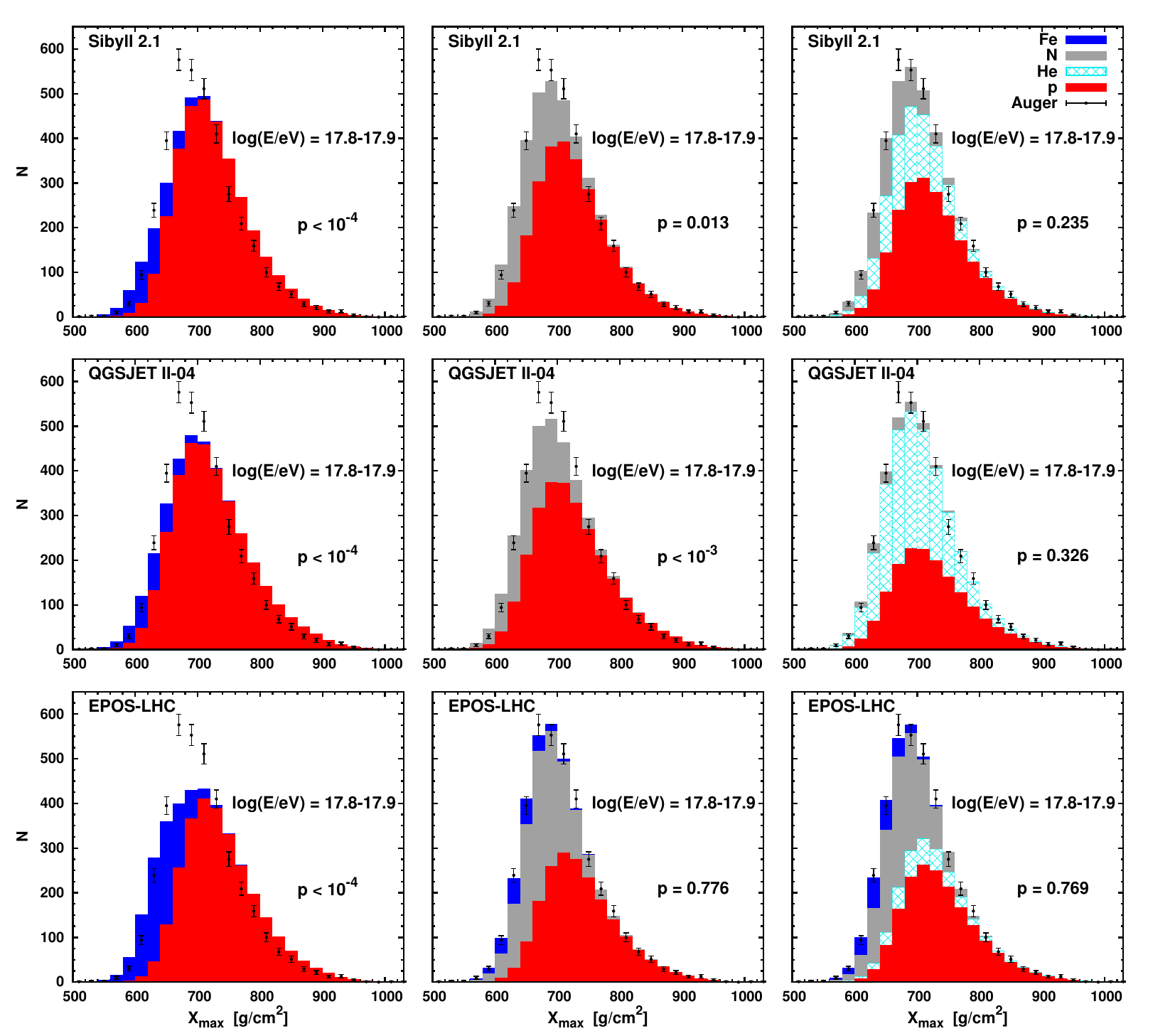}
\caption{$\xm$ distribution of the fits for energy bin $E = 10^{17.8-17.9}$ eV. Results using Sibyll 2.1 are shown in the top row, QGSJET II-4 in the middle row, and EPOS-LHC in the bottom row. The left column displays results where protons and iron nuclei were used, the central column also includes nitrogen nuclei, and the right column includes helium nuclei in addition.}
\label{fig:histos-e01}
\end{figure*}

\begin{figure*}
\includegraphics[width=0.99\textwidth]{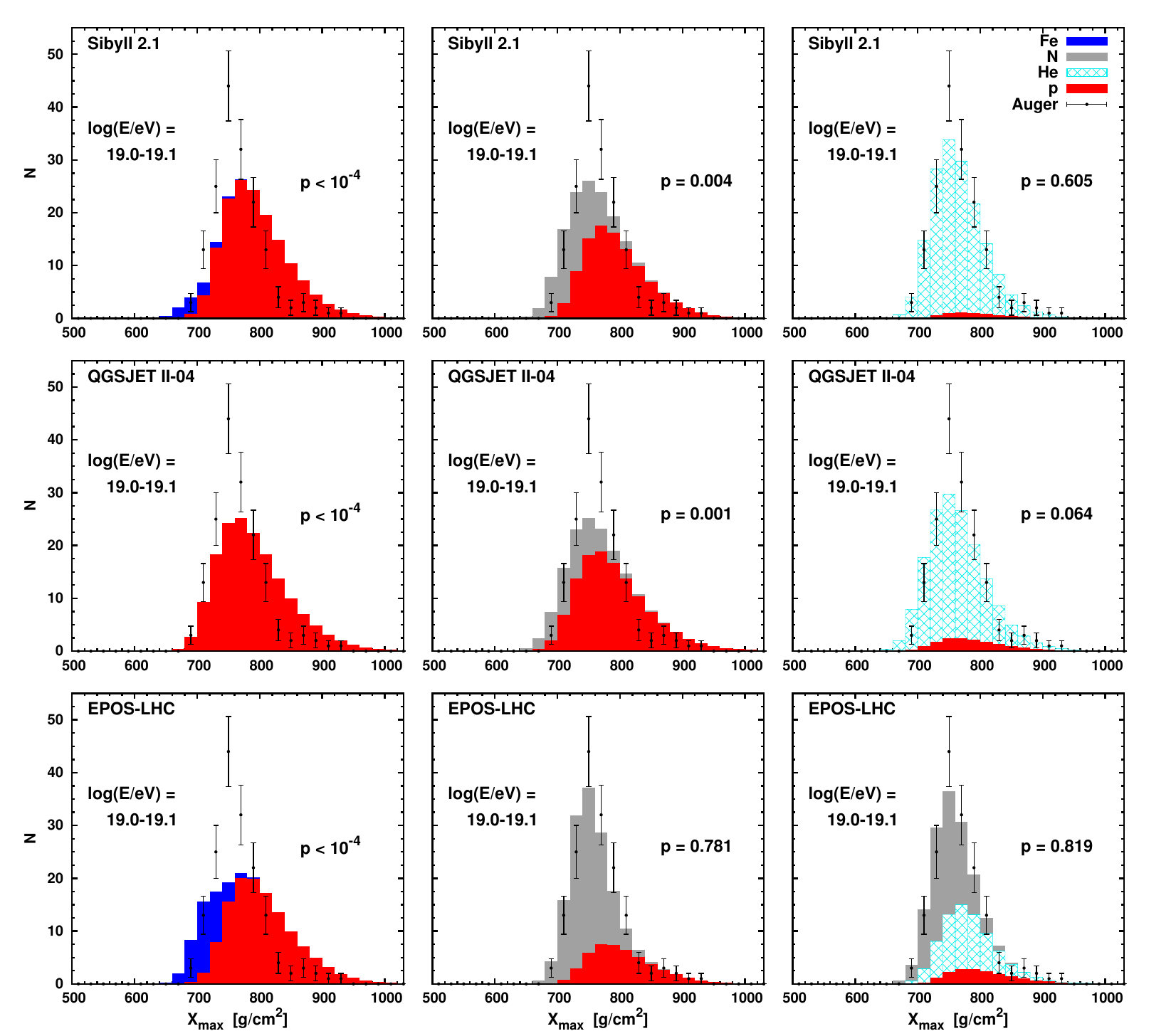}
\caption{$\xm$ distribution of the fits for energy bin $E = 10^{19.0-19.1}$ eV. See caption to Fig.~\ref{fig:histos-e01}.}
\label{fig:histos-e13}
\end{figure*}

\begin{figure*}
\includegraphics[width=0.99\textwidth]{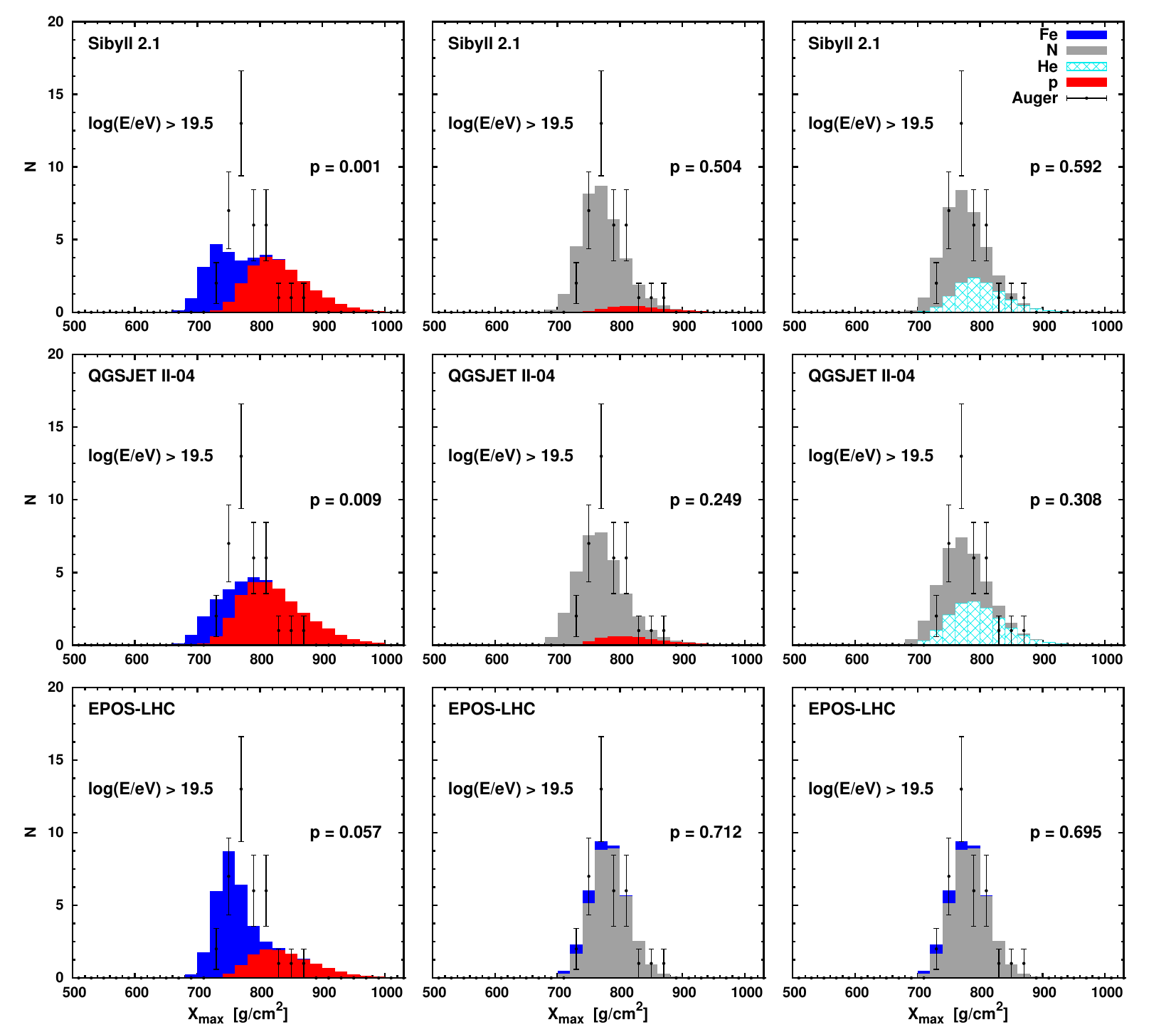}
\caption{$\xm$ distribution of the fits for energy bin $E > 10^{19.5}$ eV.  See caption to Fig.~\ref{fig:histos-e01}.}
\label{fig:histos-e18}
\end{figure*}

To aid in the discussion, the $\xm$ distributions of the fits are displayed for the energy bins $E = 10^{17.8-17.9}$ eV (Fig.~\ref{fig:histos-e01}), $E = 10^{19.0-19.1}$ eV (Fig.~\ref{fig:histos-e13}) and $E > 10^{19.5}$ (Fig.~\ref{fig:histos-e18}), respectively. Each figure contains nine panels that cover the species combination and hadronic interaction models used. The contributions of all species are stacked starting from the lightest species, with the data and their statistical uncertainty superimposed. The $p$-value of the fit is included in each panel.

\section{Discussion and conclusions}

The generally poor-quality fits obtained with the two-component scenario indicate that none of the hadronic interaction models can describe the data as a simple mixture of protons and iron nuclei. The reason for the poor fits is clear when one compares the $\xm$ distribution of the data with those predicted by the fits (see Figs.~\ref{fig:histos-e01}, \ref{fig:histos-e13}, \ref{fig:histos-e18}). The peak values for the data lie between those for protons and iron nuclei but the distributions are too narrow to accommodate a mixture of the two. Thus we conclude that either the model predictions are wrong or else other nuclei with shorter propagation length form a significant component of the UHECR flux that reaches the upper atmosphere.

Adding intermediate components greatly improves the fits for all hadronic interaction models. Results using EPOS-LHC in particular are satisfactory over most of the energy range. It is interesting to note that including intermediate components also brings the models into remarkable agreement in their predictions of the protons and iron nuclei contributions despite large differences in the remaining composition. This can be seen in the right column of Fig.~\ref{fig:histos-e01}. All three models give acceptable fit qualities with consistent fractions of protons, but with distinctly different predictions for the remaining composition; results of EPOS-LHC simulations favor a mixture dominated by nitrogen nuclei, while the QGSJET II-4 simulation favor helium nuclei, whereas Sibyll 2.1 modeling leads to a mixture of the two. 

A substantial change in the proton fractions is observed across the entire energy range, which rises to over 60\% around the ankle region ($\sim 10^{18.2}$ eV) and subsequently drops to near zero just above $10^{19}$ eV with a possible resurgence at higher energies. If the ankle feature is interpreted as a transition from Galactic to extragalactic cosmic rays \cite{Linsley:1963}, the proton fraction in this energy range is surprisingly large as the upper limits on the large-scale anisotropy \cite{Auger:2012an} suggest that protons with energies below $10^{18.5}$ eV are most likely produced by extragalactic sources. In order to accommodate a proton-dominated scenario for energies above $10^{18}$ eV \cite{Berezinsky:1988wi}, the hadronic interaction models would need to be modified considerably. The transition to heavier cosmic rays with increasing energy is reminiscent of a Peters cycle \cite{Peters:1961}, where the maximum acceleration energy of a species is proportional to its charge $Z$. However further analysis that takes into account the energy spectrum and propagation of UHECRs through the universe would be required to confirm this. Composition-sensitive data above $10^{19.5}$ eV will be needed to allow a reliable interpretation of the observed changes of composition in terms of astrophysical models (see, {\it e.g.}, Refs.~\cite{Allard:2011aa, Taylor:2011ta}).

The absence of a significant proportion of iron nuclei in the fits is easy to understand when one looks at the $\xm$ distributions for the two-component fits in Fig.~\ref{fig:histos-e18}. The $\xm$ distribution of iron nuclei is predicted in all three models to peak at substantially smaller $\xm$ than the data indicate. The widths of the data distributions do not allow much room to accommodate a significant contribution from iron nuclei.

Given that our analysis is limited in the number of species included, we cannot in general use the fit qualities as indicators of the validity of the hadronic interaction models. However, it is clear that the data can be described with EPOS-LHC even when restricted to the four species used in this analysis. Adding additional species will not change any conclusions with respect to this model. The QGSJET II-4 fit to the bin $E = 10^{19.0-19.1}$ eV, allowing in principle contributions from four species, did not in fact require any components more massive than helium nuclei. If we examine the predicted distribution (center row, right column of Fig.~\ref{fig:histos-e13}), we see that though the peak of the data distribution lines up well with that of the helium nuclei, the data distribution is too narrow to be compatible with the QGSJET II-4 prediction. Replacing the helium nuclei with a heavier species with a narrower distribution would not help the fit because its peak location would be at a value of $\xm$ that is too low, and any admixture will only exacerbate the problem with the width. Since this is generally the situation wherever QGSJET II-4 has a poor fit, we conclude that adding extra species or changing the choice of species would not help to improve the fit qualities for this model.

In conclusion, we have analyzed the distributions of depths of shower maximum measured with hybrid data from Auger and found them, using current hadronic interaction models, to be inconsistent with a composition dominated by protons, nor can they support a large contribution from iron nuclei. Introducing intermediate masses to the fits produces acceptable fit qualities for some of the hadronic interaction models used. Though the fitted compositions are in general model-dependent, all three models considered gave similar results for the evolution with energy of the proton fraction. However, it is still possible that the observed trend is not due to an evolution of composition mix, but rather to deviations from the standard extrapolations in hadronic interaction models.

\acknowledgments 

The successful installation, commissioning, and operation of the Pierre Auger Observatory would not have been possible without the strong commitment and effort from the technical and administrative staff in Malarg\"{u}e. 

We are very grateful to the following agencies and organizations for financial support: 
Comisi\'{o}n Nacional de Energ\'{\i}a At\'{o}mica, Fundaci\'{o}n Antorchas, Gobierno De La Provincia de Mendoza, Municipalidad de Malarg\"{u}e, NDM Holdings and Valle Las Le\~{n}as, in gratitude for their continuing cooperation over land access, Argentina; the Australian Research Council; Conselho Nacional de Desenvolvimento Cient\'{\i}fico e Tecnol\'{o}gico (CNPq), Financiadora de Estudos e Projetos (FINEP), Funda\c{c}\~{a}o de Amparo \`{a} Pesquisa do Estado de Rio de Janeiro (FAPERJ), S\~{a}o Paulo Research Foundation (FAPESP) Grants \# 2010/07359-6, \# 1999/05404-3, Minist\'{e}rio de Ci\^{e}ncia e Tecnologia (MCT), Brazil; MSMT-CR LG13007, 7AMB14AR005, CZ.1.05/2.1.00/03.0058 and the Czech Science Foundation grant 14-17501S, Czech Republic;  Centre de Calcul IN2P3/CNRS, Centre National de la Recherche Scientifique (CNRS), Conseil R\'{e}gional Ile-de-France, D\'{e}partement Physique Nucl\'{e}aire et Corpusculaire (PNC-IN2P3/CNRS), D\'{e}partement Sciences de l'Univers (SDU-INSU/CNRS), Institut Lagrange de Paris, ILP LABEX ANR-10-LABX-63, within the Investissements d'Avenir Programme  ANR-11-IDEX-0004-02, France; Bundesministerium f\"{u}r Bildung und Forschung (BMBF), Deutsche Forschungsgemeinschaft (DFG), Finanzministerium Baden-W\"{u}rttemberg, Helmholtz-Gemeinschaft Deutscher Forschungszentren (HGF), Ministerium f\"{u}r Wissenschaft und Forschung, Nordrhein Westfalen, Ministerium f\"{u}r Wissenschaft, Forschung und Kunst, Baden-W\"{u}rttemberg, Germany; Istituto Nazionale di Fisica Nucleare (INFN), Ministero dell'Istruzione, dell'Universit\`{a} e della Ricerca (MIUR), Gran Sasso Center for Astroparticle Physics (CFA), CETEMPS Center of Excellence, Italy; Consejo Nacional de Ciencia y Tecnolog\'{\i}a (CONACYT), Mexico; Ministerie van Onderwijs, Cultuur en Wetenschap, Nederlandse Organisatie voor Wetenschappelijk Onderzoek (NWO), Stichting voor Fundamenteel Onderzoek der Materie (FOM), Netherlands; National Centre for Research and Development, Grant Nos.ERA-NET-ASPERA/01/11 and ERA-NET-ASPERA/02/11, National Science Centre, Grant Nos. 2013/08/M/ST9/00322 and 2013/08/M/ST9/00728, Poland; Portuguese national funds and FEDER funds within COMPETE - Programa Operacional Factores de Competitividade through Funda\c{c}\~{a}o para a Ci\^{e}ncia e a Tecnologia, Portugal; Romanian Authority for Scientific Research ANCS, CNDI-UEFISCDI partnership projects nr.20/2012 and nr.194/2012, project nr.1/ASPERA2/2012 ERA-NET, PN-II-RU-PD-2011-3-0145-17, and PN-II-RU-PD-2011-3-0062, the Minister of National  Education, Programme for research - Space Technology and Advanced Research - STAR, project number 83/2013, Romania; Slovenian Research Agency, Slovenia; Comunidad de Madrid, FEDER funds, Ministerio de Educaci\'{o}n y Ciencia, Xunta de Galicia, European Community 7th Framework Program, Grant No. FP7-PEOPLE-2012-IEF-328826, Spain; The Leverhulme Foundation, Science and Technology Facilities Council, United Kingdom; Department of Energy, Contract No. DE-AC02-07CH11359, DE-FR02-04ER41300, and DE-FG02-99ER41107, National Science Foundation, Grant No. 0450696, The Grainger Foundation, USA; NAFOSTED, Vietnam; Marie Curie-IRSES/EPLANET, European Particle Physics Latin American Network, European Union 7th Framework Program, Grant No. PIRSES-2009-GA-246806; and UNESCO.


\end{document}